# Plasma-Electric Field Controlled Growth of Oriented Graphene for Energy Storage Applications

Subrata Ghosh [1,2]*, S. R. Polaki [1]*, M. Kamruddin [1], Sang Mun Jeong [2], and Kostya (Ken) Ostrikov [3,4]*

[1] Surface and Nanoscience Division, Materials Science Group, Indira Gandhi Centre for Atomic Research, Homi Bhabha National Institute, Kalpakkam, Tamil Nadu - 603102, India
[2] Department of Chemical Engineering, Chungbuk National University, 1 Chungdae-ro, Seowon-Gu, Cheongju, Chungbuk 28644, Republic of Korea
[3] School of Chemistry, Physics and Mechanical Engineering, Queensland University of Technology, Brisbane QLD 4000, Australia
[4] CSIRO-QUT Joint Sustainable Processes and Devices Laboratory, Lindfield NSW 2070, Australia

**ABSTRACT**
Graphene is well known to grow as flat sheets aligned with the growth substrate. Oriented graphene structures typically normal to the substrate have recently gained major attention. Most often, normal orientation is achieved in plasma-assisted growth and is believed to be due to plasma induced *in-built* electric field, which is usually oriented normal to the substrate. This work focuses on the effect of *in-built* electric field on growth direction, morphology, inter-connectedness, and physical properties of various configurations of graphene structures and reveals the unique dependence of these features on electric field orientation. It is shown that tilting of growth substrates from parallel to normal direction with respect to the direction of in-built plasma electric field leads to the morphological transitions from flat graphene structure, to oriented individual graphene sheets and then interconnected three-dimensional networks of oriented graphene sheets. The revealed transition of the growth orientation leads to change in wetting nature, types of defect in graphitic structures as well as affects their charge storage capacity when used as supercapacitor electrodes. This simple and versatile approach opens new opportunities for the production of potentially large batches of differently oriented and structured graphene sheets in one production run.

Corresponding authors E-mail: subrata.ghoshk@rediffmail.com; polaki@igcar.gov.in; kostya.ostrikov@qut.edu.au

## 1. INTRODUCTION

Graphene and its derivatives opens up a new window for academic as well as industrial research, due to their remarkable physical and chemical properties.[1-5] The recent trend is to transform graphene structures from two-dimensional to three dimensional, often referred in the literature as carbon nanowalls, few-layered graphene nanoflakes, graphitic petals and vertical graphene nanosheets (VGN).[6-9] It is a three-dimensional interconnected network of few layer graphene sheets standing normal to the substrate. These 3D structures opens up new ways to explore the effects of the surfaces, reactive edges, substantially increased surface-to-volume-ratio and reactivity compared to substrate-supported planar graphenes.[1, 6] Along with the intriguing properties of graphene, the most interesting characteristics of this structure include excellent mechanical[10], optical[11], thermal[12], electrical[13] and magnetic properties[8], long-term stability[14], bio-compatibility[15], capability of easy functionalization[16] and several others. The self-organized VGN with aforementioned properties emerged as an attractive advanced functional material for electron field emission, energy storage and other applications.[15-20] To realize their potential in nanoscale devices, substantial efforts have been made and still continuing to architect the VGN structures with high yield, optimized inter-sheet spacing and desired properties.

From the emerging demand for VGN synthesis, plasma enhanced chemical vapor deposition (PECVD) technique has been regarded as an effective and widely used method. However, understanding the science behind the plasma assisted growth with deterministic morphology and property is essential according to technological and scientific view point. The *in-built* electric field is believed to be the key factor for the growth of carbon nanostructures in vertical orientation.[9, 21-24] Despite the immense advances in the field, very less attention has been paid to investigate the dependency of growth on plasma induced *in-built* electric field.[25-27] The straight and aligned VGN with regular inter-sheet spacing were obtained on the substrate perpendicular to the electrode,[25] while similar morphological vertical graphene on Au/AlO$_x$/Si but not on bare Si and Cu substrate are observed in Ref. [26]. The alignment mechanism was attributed to the surface plasmon effects.[26]

On the other hand, the growth direction of graphene was found to be strongly polarized by the electric field direction and hence random distribution in the vertical sheets is observed while the substrate is placed horizontally.[28] Similar result was also reported in case of carbon nanotubes (CNT) also and explained by the torque experienced along the tube axis upon changing the substrate orientation.[29] Interestingly, Bower *et al.*[30] demonstrated the growth of CNT parallel to the surface normal of the substrate irrespective of substrate orientation and shape (flat or concaved). Apart from graphene and CNT, Lin *et al.*[31] demonstrated the change in orientation of carbon nanofiber with respect to the angle between substrate surface and *in-built* electric field. The concluding remarks from these reports are the direction of electric field influence the growth orientation. Therefore, control over the electric field direction is utmost essential to grow VGN structures with desired orientation.

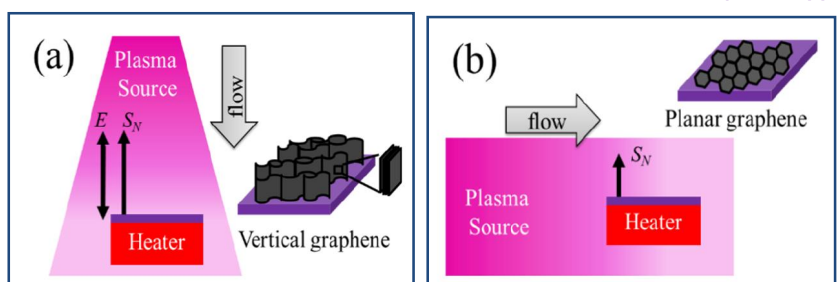

**Scheme 1:** *Schematic of PECVD for the growth of (a) Vertical graphene and (b) planar graphene*

In general, during the VGN growth in PECVD, direction of both the plasma and *in-build* electric fields are parallel to the surface normal of the substrate, as shown in **Scheme 1(a)**. Importantly, because of the relatively high rates of production and delivery of diverse building units in a plasma, it is quite challenging to achieve thin planar graphene layers using PECVD technique.[32, 33] However, recent reports on planar graphene growth in PECVD, while keeping the substrate such that plasma direction / electric field direction was perpendicular to the substrate surface normal, as depicted in **Scheme 1(b)**. [32, 33] It is also possible to grow planar graphene structure in former arrangement by choosing appropriate gas composition, substrate temperature, plasma power and deposition time.[7, 34, 35]   However, both the vertical and planar graphene structures have profound aspects depending upon the kind of application. For example, VGN are envisioned material for energy storage and field emission applications[6] whereas, Planar Nano graphitic (PNG) films attracted towards charge trapping memory.[36]

The existing background art and the many unanswered questions related to the growth process of VGN, make us pose the following questions and attempt to clarify them in this dedicated study
(i) What is the role of *in-built* electric field in controlling the growth orientation and is the normal graphene sheet orientation to the growth substrate primarily due to *in-built* electric field?
(ii) How does the direction of electric field affect the orientation, morphology, and structure of graphenes grown under the same other process conditions?
(iii) How do the arising variation in structural, morphological, wetting properties and defect presence affect performance of such graphene materials as electrodes for advanced energy storage devices?

Focusing to address these critical questions, we endeavor the growth of graphene on $SiO_2$/Si substrate placed at different inclinations, simultaneously, in PECVD system. The morphology and structural investigations were carried out. This study evidenced the significant role of plasma induced *in-built* electric field by flipping the growth from vertical to planar graphene by just inclining the surface normal of the substrate from 0 to 90º. The influence of growth orientation on their performance as supercapacitor electrodes is demonstrated. Herein, we envisaged the possibility of tuning the growth orientation of graphene structures towards extending its potential utilization over wide range of applications.

## 2. EXPERIMENTAL METHODOLOGY
### 2.1. Growth of graphene
Towards addressing the issue of *in-built* electric field orientation growth in plasma, we carried out growth of VGN on substrates with different inclinations, simultaneously. To achieve this, we designed a multi-angled faceted Cu block with surfaces at different angle of inclination ($\theta$) namely 0, 30, 45, 60 and 90°, as depicted in **Figure 1(a)**. The angle of inclination ($\theta$) is defined as angle between surface normal of the substrate and plasma induced inherent electric field [**Figure 1(a)**]. Thermally oxidized Si is used as substrate and it is placed on the Cu block on each edge. The block was placed on the heater stage and the substrates are tightened by screws. We grow VGN structures on $SiO_2$/Si substrate using Electron-Cyclotron Resonance (ECR)-PECVD technique and the details of the growth parameters were described in our previous report.[7] In brief, the deposition chamber was evacuated to $10^{-6}$ mbar, using turbo molecular pump backed by a rotary pump. Thereafter, the substrate was heated for four hours before starting the growth at 800 °C. It is anticipated that all surfaces of Cu block experience uniform temperature because of its high thermal conductivity. The distance of heater and plasma ignition point was 30 cm and maintained during the process. Pre-cleaning of the substrate was carried out by purging Ar gas with 20 sccm flow rate at microwave plasma of 375 W. The deposition of carbon species was performed by feeding $CH_4$ of 5 sccm for 30 min along with Ar career gas at same temperature and microwave power. The observed operating pressure was $10^{-3}$ mbar during the growth. Plasma and gas flow were turned off and substrate are allowed to anneal for 30 min at same temperature and base pressure for better quality of the film.[37] Thereafter, substrates were allowed for cooling down to room temperature naturally and taken out of the chamber for further characterization.

**2.2. Characterization**
The surface morphology of as-grown graphene structures was inspected by Field Emission Scanning Electron Microscopy (FESEM, Supra 55, Zeiss, Germany). High-resolution Transmission Electron Microscopy (HRTEM, LIBRA 200 FE, Zeiss, Germany) was adopted in order to investigate the microstructure and the number of graphene layers in each sheet. Wetting properties of the as-grown film were studied by releasing a water drop with volume 1 µl and the corresponding contact angle was measured (Apex instrument Co. Pvt. Ltd., India). The morphological analysis in terms of density, height and length of vertically oriented graphene sheets as well as water contact angle of as-grown films were estimated by ImageJ software. The degree of graphitization of the film was investigated by Raman Spectroscopy (*in-Via* Renishaw, UK). Raman Spectra of the films were collected using 532 nm excitation, 50× objective lens and monochromator with grating of 1800 grooves/mm. The areal capacitance of the graphene structures grown at different angle of inclinations was estimated by cyclic voltammetry. The cyclic voltammogram were carried out in a three-electrode configuration using Metrohm Autolab workstation (model PGSTAT302N, Netherland). A 1×1 $cm^2$ area of film used as working electrodes, whereas platinum foil and Ag/AgCl (KCl saturated) were used as counter and reference electrodes, respectively. Aqueous $H_2SO_4$ solution was used as electrolyte. The areal capacitance (C in F/$cm^2$) was estimated from cyclic voltammogram using the **Equation 1**: $C_s = (\int I\, dV)/v.s$ ; where *I*, *V*, *v* and *s* are current density, potential window, scan rate and exposed area of electrode, respectively.

**3. RESULTS**
**3.1. Morphological analysis**
The FESEM micrographs of the graphitic structures are illustrated in **Figure 1 (b-f)**. A flipping in morphology from three-dimensional vertical graphene to the planar structure with increase angle of inclination (0 to 90°) is evidenced. The growth of interconnected and vertically aligned three-dimensional graphene sheets network is found while the surface normal of the substrate and electric field are parallel, as shown in **Figure 1(b)**. In contrast, the PNG film with a very small number of vertically oriented flakes is obtained in the perpendicular configuration, as shown in **Figure 1(f)**. The observed results clearly indicating a significant influence of angle of inclination on the growth orientation. This result is in good agreement with the observation of existing literature on plasma induced tuning of the growth orientation in case of carbon fiber and carbon tubes.[30, 31] Here, we assumed that the top surface of Cu block experienced the same temperature irrespective of the angle of inclination. Since, the block is only 1 cm thickness and kept at 800 $^0$C for 4 hrs before starting the growth to achieve uniform temperature. We also excluded the affect from the height difference. However, to avoid the ambiguity, we reconfirmed our findings by growing the film on the Cu block (0º) and directly on the heater surface simultaneously. For the comparison sake, we have shown the morphology of VGN grown on the top (0º) of Cu block and VGN grown directly on the heater surface (0º) in **Figure 1(b) and 1(g)**, respectively. There is no significant change in the morphology is observed, except a slight variation in areal density. We also investigated the growth of VGN with respect to the distance between the plasma source and the substrate surface in our previous report.[7] Hence, the present observation evidenced the influence of *in-built* plasma induced electric field on VGN growth and possible explanation is discussed in the subsequent paragraphs.

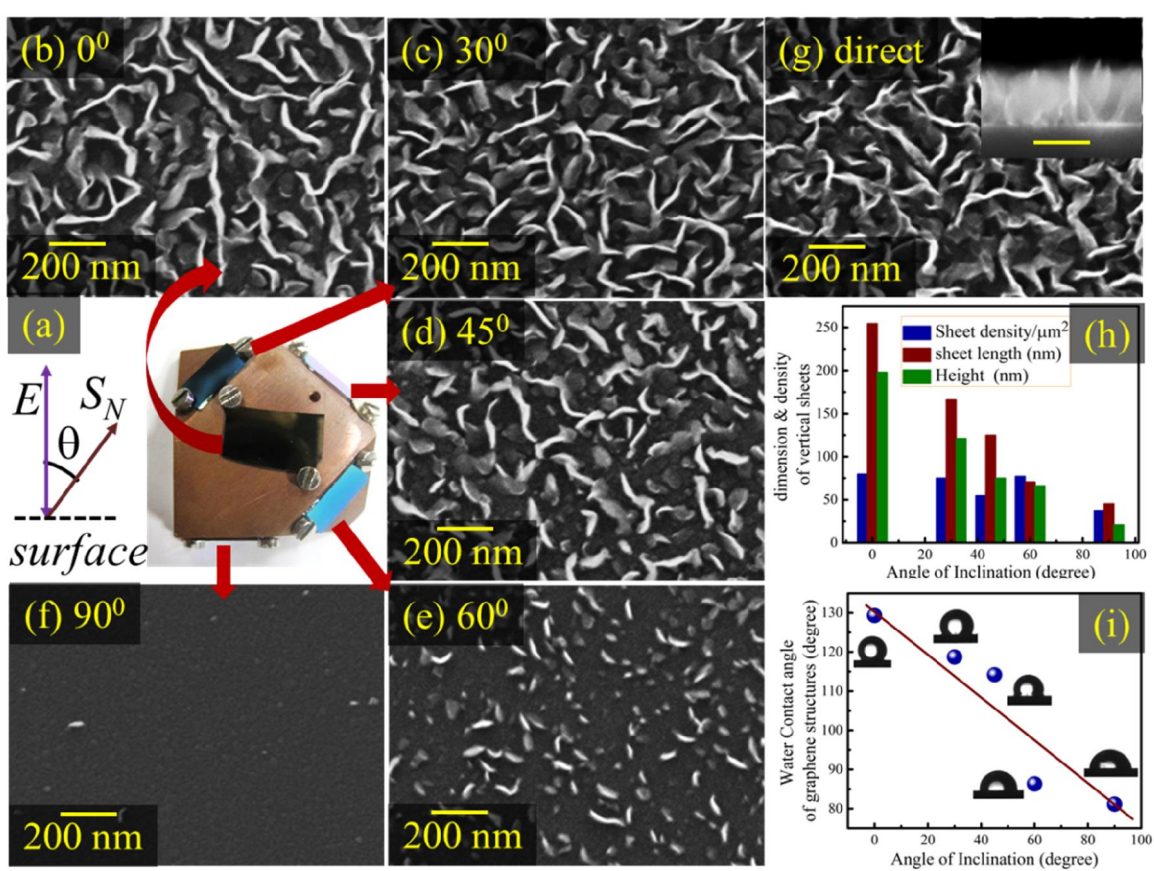

*Figure 1: (a) Photographic image of designed Cu block with different angle of inclination from 0 to 90º. Angle is defined between in-built electric field or plasma direction and normal of the substrate surface. (b-f) Scanning electron micrograph of graphene structure grown at different angle with respect to the plasma direction. (g) Micrograph of vertical graphene grown directly on the heater surface and inset represent its' cross sectional view. (h) Average vertical sheet density, average length and average height of vertical sheets and (i) water contact angle for the film grown at different angle of inclination.*

The FESEM micrographs (**Figure 1**) demonstrated the possibility of growing interconnected VGN, free standing VGN and PNG in a single growth process. As the angle of inclination reduces from 90º to 0º, the observed variations in morphology are: (i) flipping the growth from PNG to VGN, (ii) more nucleation of vertical sheets, (iii) increase in height of the sheets and (iv) merging of vertical sheets resulted in longer vertical sheets. The vertical height of the VGN is examined by cross-sectional micrograph, as shown in inset of **Figure 1(g)**. Average sheet density, average length and height of vertical sheets for the film grown at different inclinations are plotted in **Figure 1(h)**. The morphological inspection of as-grown films is also carried out via water contact angle measurement. It has seen that the contact angle of the film drops down from 129º to 81º as the film changed its orientation from vertical to planar. Hence, the change in wettability of the films is attributed to the variation in surface morphology [**Figure 1(i)**].

**Figures 2(a-b)** and **2(c-d)** depict the HRTEM micrographs of both PNG and VGN samples, respectively. It is clearly observable that the NG layer in PNG consists of around 40 number of graphene layers, whereas it is 3-10 layers of graphene layers in case of VGN structures.

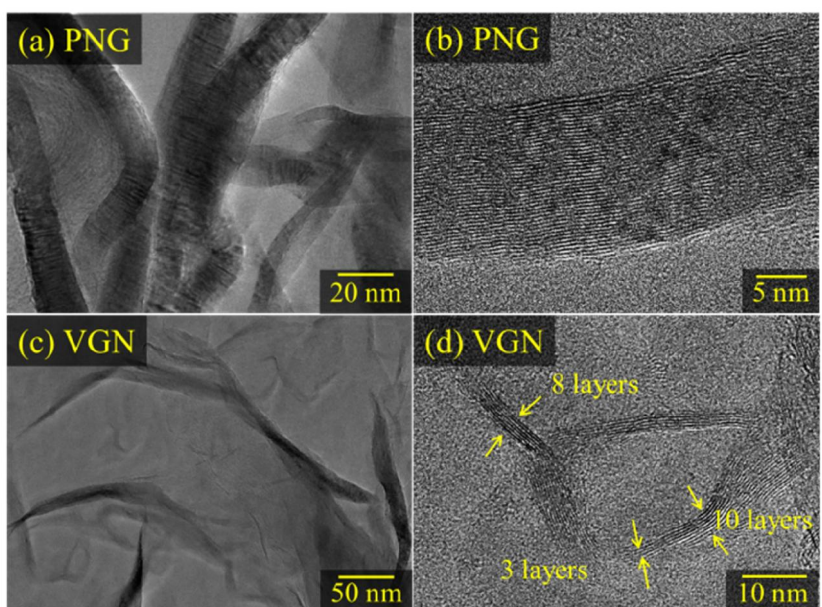

*Figure 2:* High resolution transmission electron micrograph of (a-b) planar nanographite (grown at inclination angle of 90 º) and (c-d) vertical graphene nanosheets (grown at inclination angle of 0 º) with different magnification.

### 3.2. Raman Spectra analysis
In order to study the crystallinity and chemical structure, Raman spectra are recorded in the frequency range of 1000-3500 cm$^{-1}$ for all the samples. All the Raman spectra are normalized with respect to the G- peak intensity and plotted in **Figure 3(a)**. The typical Raman spectra of PECVD grown graphene consists of D-peak at 1350 cm$^{-1}$, G peak at 1580 cm$^{-1}$ and G′ peak around 2700 cm$^{-1}$.[38, 39] The presence of G-peak is attributed to the *in-plane* stretching vibration ($E_{2g}$ symmetry) in $sp^2$ bonded carbon. The double and triple resonances are responsible for the G′-peak. The presence of G- and G′- peak affirms the graphitic nature of the grown samples [40]. On the other hand, D-band is associated with *out-of-plane* breathing mode ($A_{1g}$) of the $sp^2$ bonded carbon. The D-band along with its overtones D′ (ca. 1620 cm$^{-1}$) and D+D′ (ca.

2950 cm$^{-1}$) are attributed to the edges, point defects such as vacancies, impurities, C-H $sp^3$ bonded defects, grain boundaries, atmosphere adsorbed molecular doping and folding/bending in the structure.[41] The striking observation from the Raman spectra is the reduction in intensity of D-peak, merging the shoulder of D′ peak with G-peak and red-shift of G-peak position around 8 cm$^{-1}$, as the structure transformed from vertical to planar. Hence, the peak position and full width at half maximum (FWHM) of D-, G- and G′ peaks are extracted from the Raman spectra are plotted in **Figure 3(b)**. High intense D-peak of VGN compared to PNG is attributed to its plenty of edges. It is seen that the change in geometry from VGN to PNG resulted with broadening in FWHM of D-, G- and G′ peaks. The broadening in FWHM and red-shift of G-peak position are attributed to the defects. To identify the nature of defects, the intensity ratio of D-to-D′ (I$_D$/I$_{D′}$) of each film is estimated and plotted in **Figure 3(b)**. The I$_D$/I$_{D′}$ ratios for the VGN and PNG are found to be 5.9 and 2.8, respectively. Based on the existing literature, this measured value implies that VGN contains vacancy-like defects whereas boundary-like defects dominated in PNG. [42]

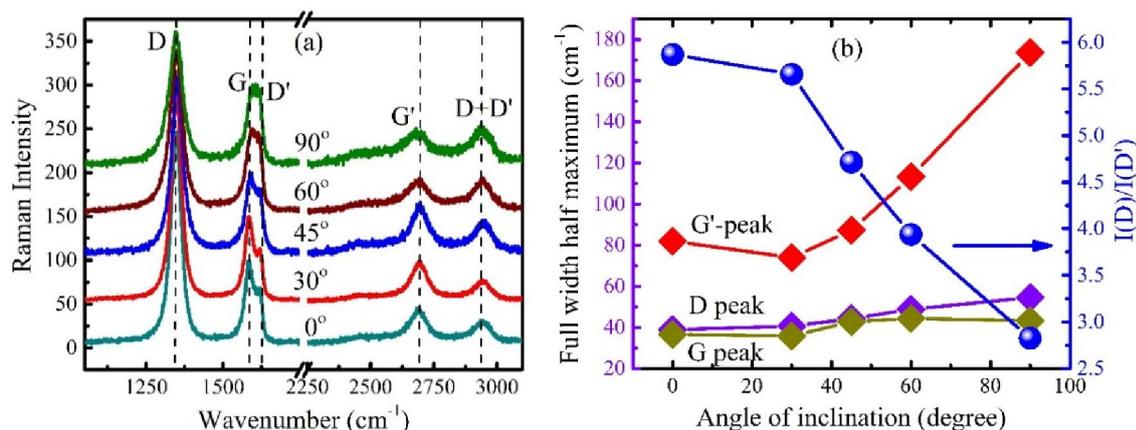

*Figure 3*: (a) Raman spectra and (b) Full width at half maximum of D-, G- and G′ peak and intensity ratio of D-to-D′(I$_D$/I$_{D′}$) for nano-graphitic film grown at different angle of inclination.

### 3.3. Performance as Supercapacitor electrodes

In order to find out the correlation between geometry and energy storage capacity, cyclic voltammetry studies were carried out on all the samples. The measured cyclic voltammograms (C-V) are depicted in **Figure 4(a)**. It has been clearly seen that as the graphene structure transformed orthogonally to the substrate with perfect alignment, the current response under C-V substantially increased. The areal capacitance of the graphene films is estimated using **Equation (1)** and plotted in **Figure 4(b)**. The areal capacitance of VGN is found to be 191 µF/cm$^2$; it is roughly eight times higher than that of the PNG structure (26 µF/cm$^2$). In general, hydrophilic graphene structures possess higher capacitance. However, the hydrophilic PNG graphene exhibited lower capacitance compared to the vertical graphene. This is attributed to the high surface to volume ratio of the vertical graphene structures. The observed results emphasize that vertical orientation of graphene structures is preferred over horizontal comes to the charge storage applications.

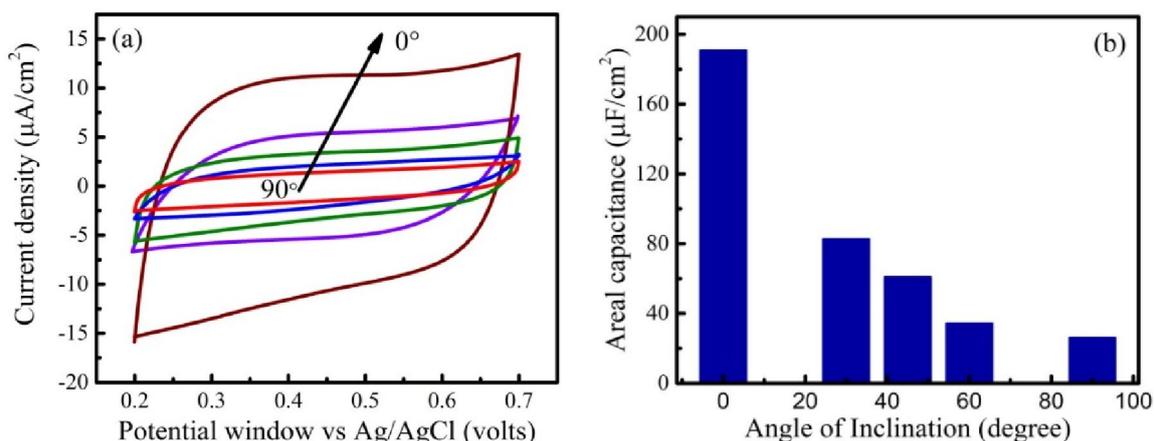

*Figure 4*: Cyclic voltammogram for the film grown at different inclination. The arrow indicates the descending order of inclination angle, (b) plot of areal capacitance versus film grown at different inclination.

## 4. DISCUSSION

The mechanism behind flipping in growth orientation of graphene structure under the influence of *in-built* electric field is as follows. It is well known that plasma the fourth state of matter consist of ions, radicals, electrons and neutrals etc. Amongst, due their lighter mass electrons are more mobile and hence the surface of all the floating materials including substrate acquires negative charge. Therefore, an *in-built* electric field is induced by the plasma onto the substrate surface. [43] Furthermore, this electric field acts as the driving force to attract the positive ions towards the substrate surface, in this case carbon ions/radicals are deposited onto the substrate. As the growth proceeds a thin PNG layer forms on the substrate in the first stage. Formation of an electrically continuous PNG within fraction of second growth is observed in our previous report. [34]

It is also worth emphasizing that the growth of graphene related structures using PECVD is the balance between depositions of carbon species and etching process by atomic hydrogen. Hence, the thickness of the PNG layer is limited. Herein, we have observed a growth of PNG layer with a few randomly nucleated vertical sheets, while the surface normal of the substrate is kept perpendicular to the *in-built* electric field direction (inclination angle is 90°). Nucleation of vertical sheets is attributed to the releasing stress through the grain boundary of NG islands as well as lattice mismatch between substrate and NG layer.

On the other hand, when the substrate surface normal is parallel to the *in-built* electric field direction (inclination angle is 0°), an upward pull by the electric field along with the stress and lattice mismatch lead to the vertical growth of graphene sheets. The strength of this force determines the alignment, ordering and growth rate of vertically oriented carbon nanostructures in PECVD. [27] The upward pull (*F*) is directly proportional to the cosine of inclination angle ($\cos\theta$) via the relation: $F \propto E \cos\theta$, E is the *in-built* electric field due to plasma. Therefore, the maximum upward pull is experienced by the horizontally placed substrate which leads to interconnected network of VGN. The strength of plasma assisted upward pulling force is reduced upon increasing the inclination angle. In the case of vertically placed substrate, the upward pull is zero hence the growth is stopped after the initial nucleation of continuous PNG films. The quantification of the electric field strength with respect to the substrate surface orientation has yet to be investigated. Since, in our previous report demonstrated that the type of substrate only has the influence on growth rate, [44] we anticipated the validity of obtained results irrespective of substrate. Hence, this fact clearly evidenced the angle of inclination of electric field has predominant role in tuning the morphology from planar to vertical graphene and hence its' graphitic quality, wettability and energy storage capacity.

## 5. CONCLUSION

In summary, we have demonstrated that the flipping in growth orientation of graphene structures by simultaneous growth on $SiO_2$/Si substrates with different inclinations achieved by a multi angle faceted Cu substrate holder in the Electron-Cyclotron Resonance (ECR) PECVD technique. The vertically aligned graphene is realized while placing the surface normal of the substrate parallel to the *in-built* electric field. In contrast, planar nano-graphitic structure is grown while the surface normal of the substrate is perpendicular to the in-built electric field. The growth rate and dimension of the sheets of VGN showed a strong dependency on the inclination angle of electric field. The dramatic change in growth orientation of graphene structures are corroborated with the strength of *in-build* electric field. A significant influence of growth orientation on the structural and wetting properties are evidenced. These features in turn affect performance of graphene sheets with different orientation and structures as electrodes in supercapacitor energy storage devices. Interestingly, the capacitance of VGN is measured to be 191 $\mu F/cm^2$, which is approximately eight times higher than that of the PNG structure (26 $\mu F/cm^2$). This study offers a simple, versatile and easily scalable process for the effective control of the morphological and structural properties of graphene sheets and layers and can be used for customized production of batches of graphenes of very diverse types and properties in the same process run.

**Notes**
The authors declare no competing financial interest.

**Author Contributions**
S. G. conceived the idea, performed the experiments, analyzed the data and wrote the draft. S. G., S. R. P. and K. O. contributed to the draft review. All authors discussed the results, commented on the manuscript and gave approval to the final version of the manuscript.

## ACKNOWLEDGEMENTS
S. G. and S. R. P. acknowledge Department of Atomic Energy, Govt. of India. S. G. also acknowledges financial support from Basic Science Research Program (2017R1D1A1B03028311) of the National Research Foundation of Korea. K. O acknowledges partial support from the Australian Research Council and CSIRO. We are grateful S. Amirthapandian for the assistance with TEM. We also acknowledge K. Ganesan and G. Amarendra for their kind support and assistance